\documentclass[final,12pt]{elsarticle}
\biboptions{sort&compress}

\usepackage{amssymb}
\usepackage{url}

\usepackage{amsmath}
\usepackage{cleveref}

\crefname{figure}{fig.}{fig.}
\Crefname{Figure}{Fig.}{Fig.}
\crefname{equation}{equation}{equation}
\Crefname{Equation}{Equation}{Equation}
\journal{Signal Processing: Image Communication}

\begin{document}

\begin{frontmatter}



\title{Texture retrieval using periodically extended and adaptive curvelets}


\author[kust]{Hasan Al-Marzouqi}
\author[gatech]{Yuting Hu}
\author[gatech]{Ghassan AlRegib}

\address[kust]{Department of Electrical and Computer Engineering, Khalifa University of Science and
Technology, AbuDhabi, United Arab Emirates}
\address[gatech]{School of Electrical and Computer Engineering, Georgia Institute of Technology, Atlanta, GA, 30332, USA}

\begin{abstract}
Image retrieval is an important problem in the area of multimedia processing. This paper presents two new curvelet-based algorithms for texture retrieval which are suitable for use in constrained-memory devices. The developed algorithms are tested on three publicly available texture datasets: CUReT, Mondial-Marmi, and STex-fabric. Our experiments confirm the effectiveness of the proposed system. Furthermore, a weighted version of the proposed retrieval algorithm is proposed, which is shown to achieve promising results in the classification of seismic activities.
\end{abstract}


\end{frontmatter}


\section{Introduction}
\label{Intro}
A large amount of visual content is constantly being generated and distributed. Productive use of such visual content demands improved procedures for indexing, analyzing, and classifying  such material. These procedures rely on the visual-content of image/video databases. Textures are important components in images, and many algorithms have been proposed in recent decades aimed at developing content-based texture similarity measures. These methods are used in classifying and retrieving images of textured materials. They can be extended for use as elements in the framework of general content-based image retrieval systems. 

Texture images exhibit a repeated pattern of visual content, and such repetitions can be captured using frequency-domain techniques. The global holistic nature of such methods makes them appropriate for texture analysis. Frequency-domain texture-retrieval algorithms typically comprise two essential components: 1) a sparsity-inducing transform that divides the spatial content of images into sets of coefficients representing unique subbands, and 2) a similarity measure that computes a numerical distance between image representations.
\begin{figure}[tb]
  \centering
  \centerline{\includegraphics [width=8cm] {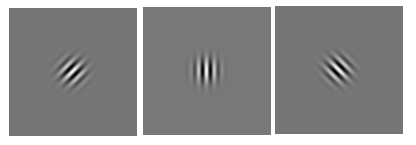}}
\caption{Three Gabor basis elements at different orientations}
\label{Gabor-elements}
\end{figure}

\begin{figure}[tb]
  \centering
  \centerline{\includegraphics [width=8cm] {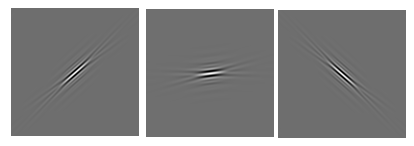}}
\caption{Examples of curvelet basis elements}
\label{Curvelet-elements}
\end{figure}
\begin{figure}[tb]
  \centering
  \centerline{\includegraphics[width=8cm]{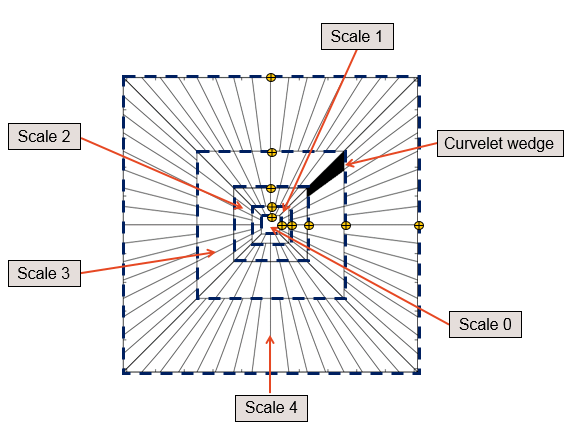}}
  \caption{ Default curvelet tiling. Scale locations are denoted by yellow markers.}
  \label{defaultcurvelet}
\end{figure}

Do and Vetterli \cite{Wavelet} proposed using the Kullback-Leibler distance between generalized Gaussian density estimates of wavelet subbands. Several variants and modifications of algorithms using wavelets for texture retrieval have been proposed in the literature \cite{Copula,Wavelet2,Wavelet3}. The wavelet transform offers limited directional selectivity (wavelet subbands contain either horizontal, vertical, or diagonal information). Directional transforms allow for more directional selectivity, and they have been used successfully in a variety of application areas. Steerable pyramids \cite{Steerable}, Gabor wavelets \cite{Gabor}, and curvelets \cite{Curvelet} are examples of directional transforms that have been used for texture retrieval. Zujovic et al. \cite{Structural} used statistical properties of a steerable pyramid representation of the texture dataset. These statistical properties included the mean, variance, horizontal and vertical autocorrelation, and cross band correlations. Gabor wavelets incorporate directional selectivity by using sets of Gaussian-shaped filters at different rotations. Manjunath et al. \cite{Gabor-texture} used the mean and standard deviation of Gabor coefficients for texture retrieval. This Gabor-based algorithm is included in the multimedia content description standard MPEG-7 \cite{MPEG-7}. Gabor wavelets decomposes the input image into a set of Gabor elements. \Cref{Gabor-elements} shows exemplary Gabor elements representing three different orientations. Zhang et al. reported performance improvements over Gabor filters by using the default curvelet transform \cite{curvelet-texture}. Curvelet elements are elongated needle-shaped elements that are generated by taking the inverse Fourier transform of anisotropic frequency bands, which we refer to as curvelet wedges. The length of each curvelet wedge is constructed to equal the square of its width. \Cref{Curvelet-elements} illustrates a number of curvelet basis elements. Default curvelet tiling of the frequency domain is shown in \Cref{defaultcurvelet}.

Several authors~\cite{LBP,DOT1,DOT2,guo2010completed,liu2015median, hu2016completed,hu2017scale,LTP, CLBC, SSLBP,long2018comparative} have developed texture-retrieval algorithms based on spatial domain content analysis. Ojala et al. \cite{LBP} developed local binary patterns (LBP) that decompose input images into sets of coefficients representing intensity differences between a reference pixel and its neighbors. Dominant orientation templates \cite{DOT1, DOT2} use a set of dominant gradient orientations to extract features representative of the object of interest. 

Guo et al. proposed completed local binary patterns (CLBP)\cite{guo2010completed} that encode both the signs and the magnitudes of local intensity differences into binary codes and combine these two complementary pieces of information using joint or hybrid feature distributions. To capture both micro- and macro-structures of images and reduce the effect of noise, Liu et al. developed median robust extended local binary patterns (MRELBP)\cite{liu2015median} using a multi-scale sampling strategy and median intensity values. This process is more robust to illumination variations, rotation variations, and noise. A comprehensive review of LBP variants is presented in \cite{liu2017local}.

More recently, deep neural networks have been used in texture retrieval \cite{cimpoi2015deep}. The proposed method is intensive in terms of both memory and storage demands. The neural network architecture used in \cite{cimpoi2015deep} employs approximately 60 million parameters, making it difficult to implement on  systems with limited hardware resources. Another interesting line of research integrates visual image features and click data from search engines to  extract relevant images \cite{yu2015learning,yu2017deep}. 

In this paper, we propose new texture-retrieval algorithms based on periodically extended and adaptive curvelets. Periodic extension improves default curvelet performance by connecting boundary elements. The adaptive curvelet \cite{al2017curvelet,aldiss} is a recently developed transform that adapts the size and location of curvelet tiles to represent a given input image more effectively. This approach adapts the direction and thickness of curvelet basis elements to  better represent image features.

Some of the results shown in this paper have appeared in \cite{CurveletGlobalSIP14}. In this paper, we introduce a rotation-normalized variant of the published algorithm, present a weighted version of the algorithm exploiting curvelet wedge properties, and expand the experimental section with experiments covering additional texture datasets. The remainder of the paper is organized as follows. In the next two sections, we will present an overview of periodically-extended and adaptive curvelet transforms, followed by technical details of the texture-retrieval algorithm in Sections 4 and 5. Experimental results are presented in Section 6. Finally, Section 7 presents our conclusions.

\section{Default curvelets with periodic extension}
The curvelet transform \cite{Curvelet} provides an efficient representation of directional features. It works by dividing the spatial content of images into different frequency bands representing unique scale and directional features. 

Curvelet coefficients are generated using the 2-D fast Fourier transform (FFT). FFT is first applied to an image; then, a scale-selection algorithm is employed to compute the optimal number of scales. Scale locations, which we define as coordinates of the FFT plane that determine the outer boundary of each scale, are determined in a dyadic manner. Each curvelet scale is further divided into a number of different directional tiles, as shown in \Cref{defaultcurvelet}.

\begin{figure}[tb]
  \centering
  \centerline{\includegraphics[width=7cm]{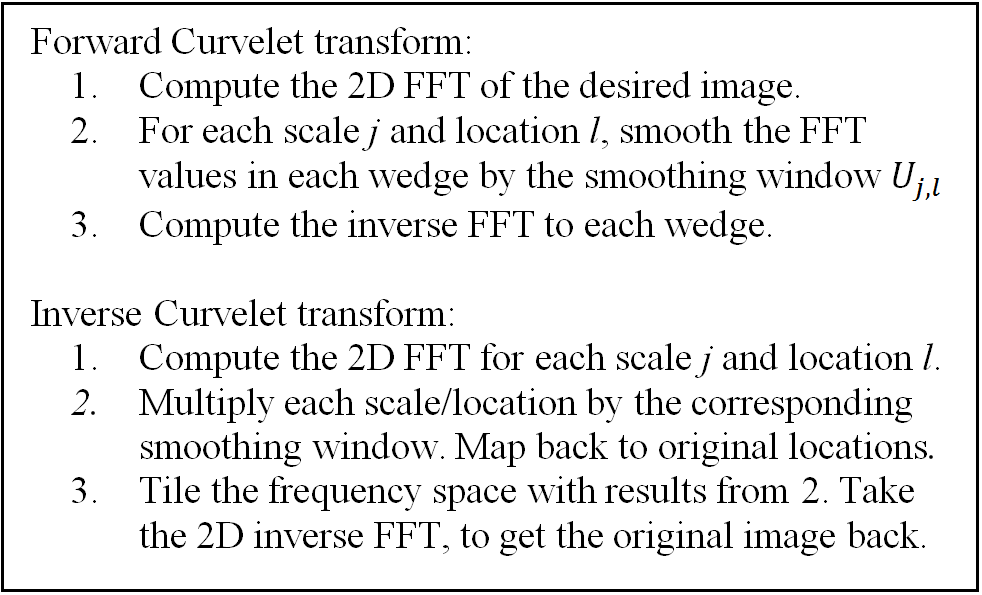}}
  \caption{Summary of the forward and inverse curvelet transforms}
  \label{curveletflow}
\end{figure}

Because FFT values for real data are symmetric around the center of the FFT plane, two quadrants are sufficient for constructing the curvelet representation. To obtain real-valued coefficients for such datasets, the complex coefficients are separated into two parts, leaving the total number of tiles unchanged. Curvelet coefficients are generated by taking the inverse FFT's of these tiles. The inverse curvelet transform restores the original image by reversing the forward transform operations. The flow of the forward and inverse curvelet algorithms is summarized in \Cref{curveletflow}

The outer curvelet scale can be constructed using either a regular high-pass filter, or by employing a periodic extension window. The first option reduces computational and storage costs, while the second option reduces boundary artifacts and increases the sparsity of the transform. A performance comparison between periodically extended curvelets and default curvelets will be presented in Section 6. Curvelets constructed with outer angular divisions and a periodic extension window are used in the proposed texture-retrieval algorithms.

\section{Adaptive Curvelets}
Adaptive curvelets alter the number, size, and locations of curvelet tiles according to a given cost function \cite{al2017curvelet}. The frequency-domain tiling that generates the best improvement in the cost function value is considered optimal. In denoising-based adaptive curvelets, the cost function is the denoising performance, as measured by the peak signal-to-noise ratio (PSNR). A synthetic noisy image $\widetilde{I}$ is generated by adding Gaussian noise with standard deviation $\sigma$ to the input image $\textit{I}$. In this work, $\sigma$ is given by
\begin{equation}\label{sigma}
\sigma=\max(\textrm{\textit{std(I)}},0.05\times \textrm{\textit{MAXI}}),
\end{equation}
where $std(I)$ is the standard deviation of the intensities of image $I$, and $\textrm{\textit{MAXI}}$ is the maximum possible intensity value. Equation (1) provides a noise level that has obvious effects on image quality, but does not overpower image details. More details regarding the relationship between $\sigma$ values and the performance of adaptive curvelets are provided in \cite{aldiss}. Next, a value for the number decomposition scales is chosen, followed by search algorithms finding the optimal scale locations and number of angular divisions for each scale/quadrant pair. The algorithms converge, returning the optimal curvelet tiling for the given image. The optimization problem characterizing denoising-based adaptive curvelets can be described mathematically by
\begin{equation}\label{denoisingbased}
\arg\max_{S,A} PSNR (curveletdenoise_{S,A}(\widetilde{I}),I),
\end{equation}
where $I$ is the input image, $\widetilde{I}$ is the artificially generated noise image, $S$ is a vector describing adaptive curvelet scale locations, $A$ is a vector that determines the number of angular decompositions used in each curvelet scale/quadrant pair, $curveletdenoise_{S,A}$ is a function that denoises the noisy image with curvelets described by $S$ and $A$, and $PSNR(A,B)$ is a function that computes the PSNR between images $A$ and $B$. The algorithms used to search for a solution for \cref{denoisingbased} are described in the next section.

\subsection{The scale-selection algorithm}
\begin{figure}[htb]
  \centering
  \centerline{\includegraphics[width=6cm]{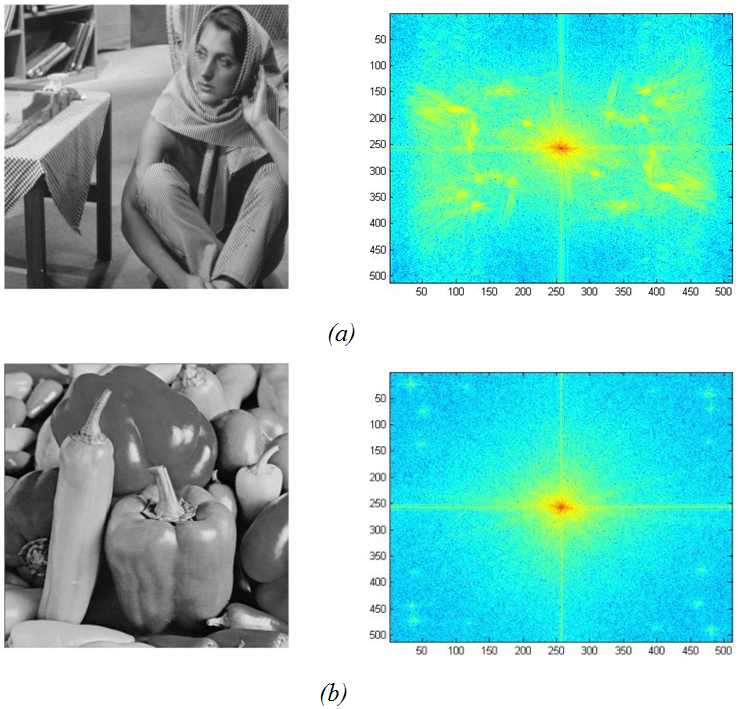}}
  \caption{ \texttt{Barbara} and \texttt{Peppers} images and their corresponding FFT log magnitude plots. FFT values decrease as color changes from red to blue.}
  \label{highfrequencycentered}
\end{figure}

Scanning a large variety of widely used images, a connected region of high magnitude FFT values can be observed in the center of the FFT plane in many of these images. An example of such activity is shown in \Cref{highfrequencycentered}, which shows a plot of FFT log magnitude for two images. It is preferable to avoid dividing such connected regions into different tiles while applying the curvelet transform, to achieve a better representation of the signal of interest, We developed a heuristic algorithm to
determine the number of scales guided by this observation in \cite{CurveletSPIE}. The algorithm works as follows:
\begin{description}
  \item [Step 1] Define a quantity termed mid-range $\mathrm{mR}$, as follows:
  \begin{equation}\label{Dvalue}
  	\mathrm{mR} \triangleq \sqrt{\max¡ \vert \mathrm{FFT value} \vert \times\min \vert \mathrm{FFT value}\vert}.
  \end{equation}
  If the minimum Fourier magnitude value is equal to zero, the next minimum value is chosen.

  \item [Step 2]let $\mathrm{D} = \mathrm{D_{initial}}$ = 8. This is the smallest possible value for the size of the inner-most curvelet level.

  \item [Step 3] Set a square surrounding the origin of the frequency domain with a diagonal length equal to $\mathrm{D}$ pixels.

  \item [Step 4] If any FFT magnitude value within the square is smaller than the mid-range value, exit the algorithm and return $\mathrm{D}$. Otherwise, go to Step 5.

  \item [Step 5]Set $\mathrm{D} = \mathrm{D + 2}$ and return to Step 3.
\end{description}
	
The optimal number of scales is found using the value $\mathrm{D}$ computed in Step 4 above. Using dyadic scaling, the optimal number of scales $\mathrm{J}$ is computed as follows:
\begin{alignat}{2}
\mathrm{J} &= \lceil(\log_{2}(\max(N_{1},N_{2})-1)-(\log_{2}(\mathrm{D})-1)\rceil \label{ssa}\\
  &= \Big \lceil\log_{2} (\frac{\max(N_{1},N_{2})}{\mathrm{D}})\Big\rceil,
\end{alignat}

where $N_{1}$ and $N_{2}$ are the vertical and horizontal image dimensions respectively. \Cref{ssa} computes the difference between the number of scales required to reach the origin of the frequency domain from the edge of the image (\textit{i.e.}, total number of scales), and the number of scales necessary to reach the origin from the outer edge of the coarsest level square. 


\subsection{Number of angular decompositions}
\begin{figure}[htb]
  \centering
  \centerline{\includegraphics[width=5cm]{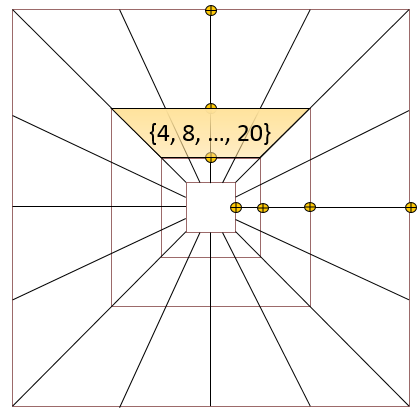}}
  \caption{Optimal angular decomposition is found through a brute-force search strategy for each scale/quadrant pair. The figure shows the second scale/first quadrant curvelet segment.}
  \label{angularscalequadrant}
\end{figure}

Optimizing the number of angular divisions is performed in a brute-force fashion in each curvelet scale/quadrant pair (\Cref{angularscalequadrant}). The tested parameters are chosen from the following sequence \{4, 8, 12, 16, 20\}. The number of divisions that achieve the maximum cost function value is considered optimal. The divisions are uniformly distributed in each scale/quadrant pair. Recalling FFT’s symmetry for real data, the number of parameters to optimize for real data is equal to $2\times(J-1)$, and is equal to $4\times(J-1)$ for complex input data. 

\subsection{Scale-locations search}
We propose using derivative-free optimization methods to search for adaptive curvelet scale-locations. Such methods start at an initial point x0, evaluate the cost function at a selected mesh of points in the neighborhood of x0, and define the point in the mesh with the minimum function value as the new x0. Next, the algorithm iterates until a specified convergence criterion is met. The Nelder-Mead simplex search method \cite{conn2009introduction} is a popular method for generating such a mesh. It has been used extensively in a variety of application areas. In this work, it is used to find the optimal scale locations. Given a specific angular distribution, the scale location search algorithm finds optimal vectors \textbf{H} and \textbf{V}, where  $\textbf{V} = \{$\textit{V$_{1}$}$, $\textit{V$_{2}$}$, ..., $\textit{V$_{J}$}$\}$ are the coordinates of vertical scale locations, and $\textbf{H} =$ \{$\textit{H$_{1}$}$, $\textit{H$_{2}$}$, ..., $\textit{H$_{J}$}$\} are the horizontal scale locations (\Cref{defaultcurvelet}). We do not constrain curvelet scales to be of the same length and width. This makes the number of optimizing parameters equal to $2\times J$ for real data.

\subsection{Global optimization algorithm}
\begin{figure}[tb]
  \centering
  \centerline{\includegraphics [width=13.5cm] {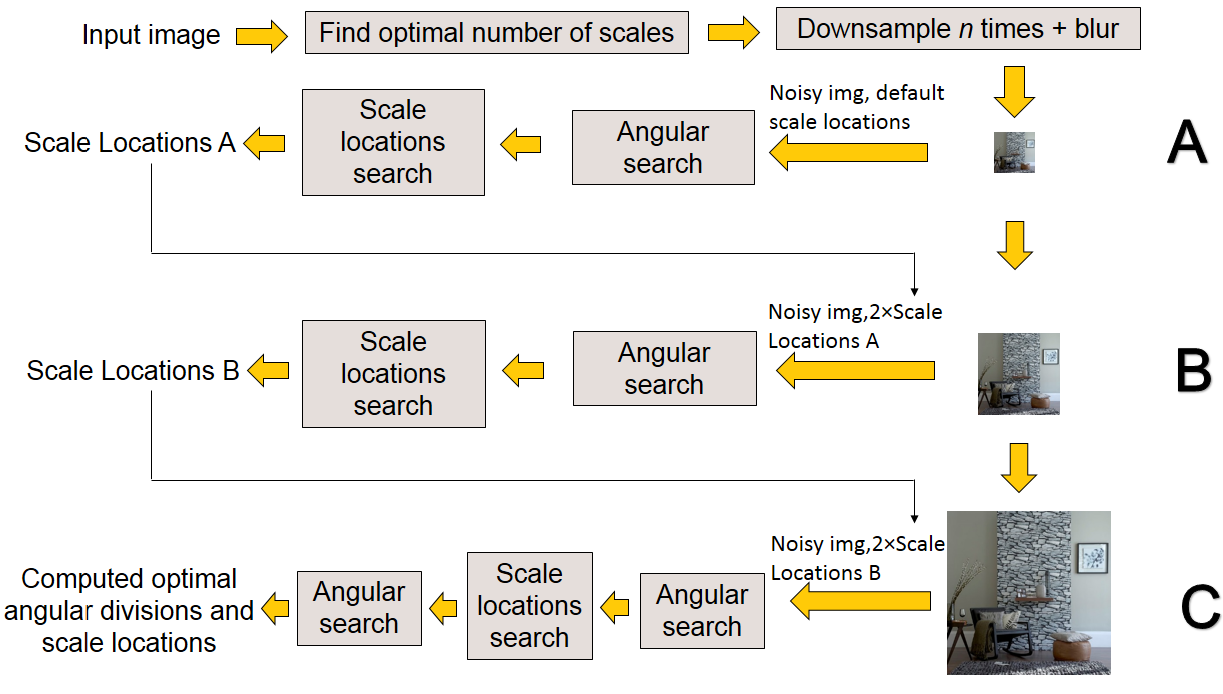}}
\caption{Flow diagram of the global optimization algorithm.}
\label{algflowchart}
\end{figure}

In this section, we introduce the global optimization algorithm that combines the two previous adaptations. The algorithm uses a multi-resolution search strategy, which helps in avoiding convergence to local maxima and reduces the computational cost of the algorithm. Optimization is conducted in a hierarchical manner consisting of $n =3$ iterations. In each iteration, optimal scale and angular locations are found for a downsampled smoothed version of the input image. Three iterations were found to be adequate for image sizes up to 512 x 512 \cite{aldiss}. Increasing the number of iterations did not generate further improvements in performance. The flow of this algorithm is illustrated in \Cref{algflowchart}.
The algorithm terminates returning the computed optimal scale and angular decompositions. The returned tiles will be used in the next section to extract feature vectors for use in texture-retrieval applications.

\section{Feature vector and distance computation}
Forming a representative feature vector is essential for the success of transform-domain retrieval methods. Using the curvelet transform, the query and input images are represented by sets of coefficients identifying the spatial content relative to unique frequency bands. Curvelet elements are represented in the spatial domain by needle-shaped elongated objects, as illustrated in \Cref{Curvelet-elements}.

The similarity between two curvelet tiles representing a certain frequency band is indicative of the similarity of the spatial content in this band between the two images. Therefore, a global similarity measure between two images can be formed by including the similarities between curvelet tiles across all frequency bands. Given that edges are essential components in images, a curvelet-based similarity measure between curvelet tiles is perceptually meaningful. Furthermore, adapting curvelet coefficients to represent the query image more effectively constructs a representation from curvelet objects that is more faithful in direction and scale to the features of the image of interest.

Given two images $A$ and $B$, where $A$ is the query image, we propose the following two algorithms for texture retrieval. The first algorithm uses default curvelets with periodic extension, while the second algorithm employs denoising-based adaptive curvelets. The periodically extended curvelet and adaptive curvelets are used to compute the curvelet coefficients representing images $A$ and $B$. In adaptive curvelets, the optimal curvelet tiling will be learned using the query image $A$. The learned frequency-domain tiling will be used to compute feature vectors representing images $A$ and $B$.

The feature vectors describing images $A$ and $B$ are formed by the mean and the standard deviation of curvelet coefficients in each curvelet tile. Let the number of tiles in the optimal representation be $K$. The feature vector representing images $A$ and $B$ is given by
\begin{eqnarray} \label{feature}
  F &=& \{\mu_{1},\sigma_{1},\mu_{2},\sigma_{2},....,\mu_{K},\sigma_{K}\},
\end{eqnarray}
where $\mu_{i}$ is the mean of the curvelet coefficients in tile $i$ and $\sigma_{i}$ is the standard deviation of the coefficients in the same tile.

Next, the distance between images A and B is computed by
\begin{equation}\label{distance}
  D(A,B) = \|F(A)-F(B)\|_{1},
\end{equation}
where $\|x\|_{1}$ is the $l_{1}$ norm of vector $x$. The feature vectors can be weighted to (de-)emphasize certain scales and orientation of interest (\textit{e.g.} coarse/fine or horizontal/vertical features). A general expression for a weighted feature vector $F$ is given by
\begin{eqnarray} \label{featureweighted}
  F &=& \{w_{1} \times \mu_{1},w_{1} \times \sigma_{1},w_{2} \times \mu_{2},w_{2} \times \sigma_{2},....,w_{K} \times \mu_{K},w_{K} \times \sigma_{K}\},
\end{eqnarray}
where ${\textbf{w}} \geq 0 $ is a weights vector of length equal to the number of curvelet tiles $K$, which is determined by the adaptive curvelet search method described previously. The value of $K$ that generates the best denoising performance is used to represent query images. This parameter will vary according to the details in a given input image.

\section{Coefficient sorting and approximate rotation invariance}
\label{rotation-invariant}
In this section, the proposed texture-retrieval algorithm is extended to handle rotations. Rotating an image rotates its FFT magnitudes \cite{RotationFFT}. The FFT content described by certain curvelet tiles in an image, will be moved to another curvelet tile in the curvelet representation of the rotated image. By sorting curvelet tiles in each scale, one can reduce rotation artifacts in the distance computed between feature vectors representing two images. Curvelet tiles in each scale are sorted based on the sum of their coefficients magnitude. Once curvelet tiles are sorted, the distance between the images of interest can be computed. This distance will be based on feature vectors with aligned entries. This sorting mechanism was used successfully in \cite{curvelet-texture}.

Rotating an FFT can generate the undesired artifact of moving high magnitude FFT data to lie across the boundary between two curvelet tiles. This artifact makes the feature vector less representative of the similarity between the image and its rotated version. To reduce the effect of such cases, its preferable to use small values for the number of angular divisions. The number of angular decompositions for each curvelet scale/quadrant pair is set to four.

\begin{figure}[tb]
  \centering
  \centerline{\includegraphics [width=8cm] {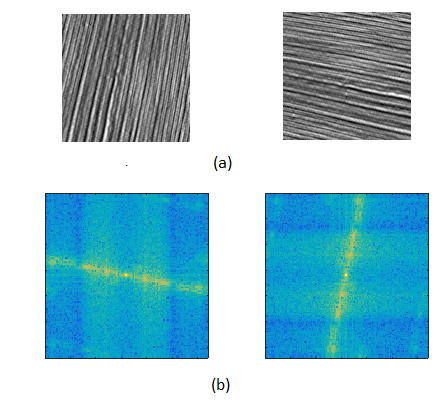}}
\caption{(a) An image and its rotation by $90^{\circ}$, (b) Corresponding log-magnitude plots of their Fourier transforms}
\label{rotation_imgs}
\end{figure}

A texture image and its $90^{\circ}$ rotated variant are shown in \Cref{rotation_imgs}. The distance $D$ between the two images using denoising-based curvelets is 388.4. Employing the rotation resilient measure, $D$ becomes 120.7.

\section{Experiments}
The proposed algorithm was tested in texture-image retrieval using several textured materials. The different qualitative measures that were used to assess the performance of the proposed algorithms are presented in this section, followed by retrieval results from the datasets used in this study. To ensure robustness to various noise realizations, adaptive curvelet results are obtained by using the average of five different trials.

\subsection{Performance metrics}
Retrieval experiments are constructed using a query image representing a class of textured material. The proposed algorithms are assigned the task of acquiring images resembling the same class of the query image from a database of testing images. The following metrics were used to evaluate the retrieval performance:
\begin{enumerate}
  \item Retrieval precision: defined as the ratio of the number of correct matches retrieved, to the total number of correct matches. Precision in this context is a function of the number of retrieved images. We will focus on precision values at one, two, and at the number of relevant images.
  \item Mean reciprocal rank (MRR): this is the mean reciprocal of the rank of the first relevant position.
  \item Mean Average Precision (MAP): this is the mean of average precision values across all queries. The average precision (AP) is given by \begin{equation}\label{AveragePrecision}
            AP = \frac{\sum_{\forall k} P(k)  }{N},
      \end{equation}
      where $k$ and $N$ are the ranks and number of relevant images, respectively.
  \item Precision-recall curves: these figures are frequently used in the analysis of retrieval performance. In such figures, precision values are plotted as a function of recall values. Recall is defined as the ratio of the number of correctly retrieved images to the number of relevant images.
\end{enumerate}

\subsection{Texture retrieval: Experimental setup}
The algorithm is tested using three different texture databases: the Columbia-Utrecht Reflectance and Texture Database (CUReT) \cite{CUReT}, Mondial Marmi collection of granite classes \cite{granite}, and the fabric dataset included in the Salzburg Texture Image Database (STex) \cite{STex}.

\begin{figure}[tb]
  \centering
  \centerline{\includegraphics[width=6cm]{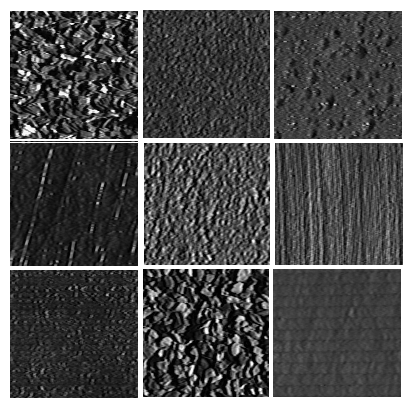}}
  \caption{A sample from the CUReT texture database}
  \label{CUReT}
\end{figure}

The CUReT database contains 61 images of real-world surfaces taken at different illumination levels and viewing directions. In our experiments, we used images taken at the first illumination setting with viewing direction 22. CUReT images are of size 640 $\times$ 480. They all include a textural part and a background. Three images of size 128 $\times$ 128 covering the textural regions were extracted from each CUReT image. One of these images was selected randomly as a query image. The remaining images were used as testing images. A sample of the images used in this study is shown in \Cref{CUReT}.

\begin{figure}[tb]
  \centering
  \centerline{\includegraphics[width=6cm]{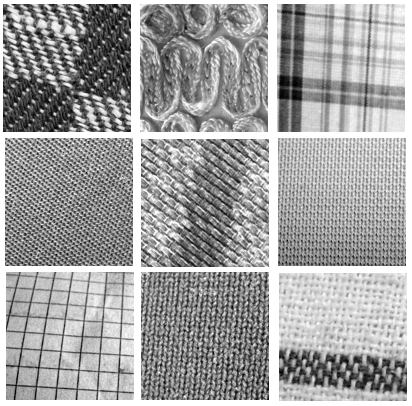}}
  \caption{Samples from STex fabric dataset}
  \label{STex}
\end{figure}

The STex dataset includes 476 images, representing different types of materials. In our experiments, we use the fabric class of images in this dataset, which includes 77 fabric samples of size 512 $\times 512$. Each image is divided into four non-overlapping regions. This makes the total number of images equal to 308. An image in each class is chosen as a query image, and the remaining images are used as testing images. A sample of the images used in this study is shown in \Cref{STex}.

\begin{figure}[tb]
  \centering
  \centerline{\includegraphics[width=6cm]{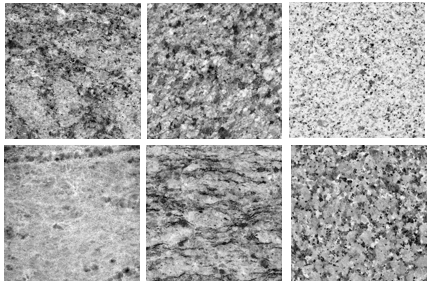}}
  \caption{A sample from Mondial Marmi texture database}
  \label{Mondial}
\end{figure}

The Mondial Marmi is a collection of 48 images representing 12 different granite classes. Each image is hardware-rotated using nine rotation angles: $0^{\circ}$, $5^{\circ}$, $10^{\circ}$, $15^{\circ}$, $30^{\circ}$, $45^{\circ}$, $60^{\circ}$, $75^{\circ}$, and $90^{\circ}$. The images were acquired under controlled illumination conditions, and each image is of size 544 $\times$ 544. To save computational time, the granite images were downsampled using a downsampling factor of 1/3. Query images included an image from each class along with its rotated versions. The remaining images were used as testing images. The number of correct matches for each query image is equal to 3 $\times$ 9. Samples from the Mondial Marmi dataset are shown in \Cref{Mondial}.

\subsection{Texture-retrieval results}
The performance of the proposed methods is compared with the following algorithms:
\begin{enumerate}
  \item $l^{2}$ norm of default curvelet features \cite{curvelet-texture}. This algorithm uses no periodic extension or angular divisions in the outer curvelet scale
  \item $l^{1}$ norm of Gabor features \cite{Gabor}
  \item Kullback-Leibler distance on wavelet features \cite{Wavelet}
  \item Log-likelihood ratio of local binary patterns \cite{LBP}. The used feature vector is composed of the concatenation of $LBP_{8,1}^{riu2}$ and $LBP_{24,3}^{riu2}$ features. In addition, we compare the performance of proposed methods with the following two variants of LBP:
  \begin{enumerate}
    \item Completed LBP(CLBP)~\cite{guo2010completed} (CLBP): We used the joint histogram of $CLBP\_C$, $CLBP\_S_{2,8}^{riu2}$, and $CLBP\_M_{2,8}^{riu2}$. 
    \item Median robust extended LBP (MRELBP)~\cite{liu2015median}: The joint histogram of $MRELBP\_CI$, $MRELBP\_NI_{2,8}^{riu2}$, and $MRELBP\_RD_{2,8}^{riu2}$ was used.
\end{enumerate}
    
\end{enumerate}

Implementations of the algorithms based on Gabor, wavelet, and LBP were downloaded from the author's websites. 


\begin{table}[tb]
\begin{center}
 \caption{Retrieval results for the CUReT dataset}
  \begin{tabular}  { |c|c|c|c|c| }
    \hline
     & P@1 & P@2 & MRR  & MAP \\ \hline
       Wavelet features &   0.967  &  0.861 &  0.980  &  0.924   \\ \hline

  Gabor features & 0.934 &0.869 &  0.962&    0.901\\ \hline
  Linear Binary Patterns (LBP) &  0.934  &   0.885    & 0.959&    0.925 \\ \hline
  Completed LBP (CLBP) &   0.984  &   0.918    & 0.986 &    0.955 \\ \hline
  Median Robust Extended LBP (MRELBP) &  0.967  &   0.869    & 0.981 &    0.924 \\ \hline
  Default curvelet &    0.951 &   0.861 &   0.975  &  0.916 \\ \hline
  Curvelet with periodic extension & \textbf{1.000} &  0.934   &\textbf{1.000} &0.964 \\ \hline
  Adaptive curvelets & \textbf{1.000}  &   \textbf{0.939}   & \textbf{1.000} &  \textbf{0.968}  \\ \hline
  \end{tabular}\label{TCUReT}
\end{center}
\end{table}

\begin{figure}[tb]
  \centering

  \centerline{\includegraphics[width=7cm]{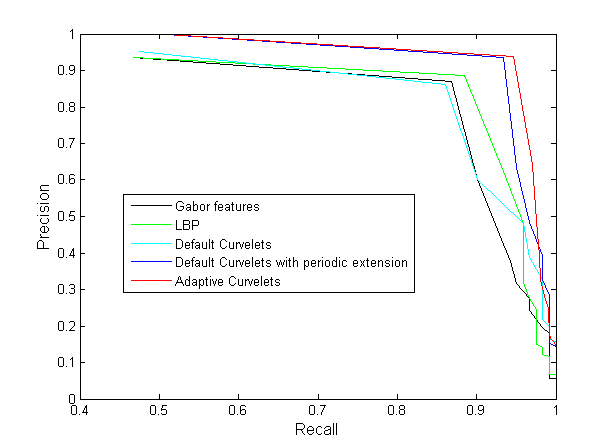}}
  \caption{CUReT Precision-recall curves for five different algorithms used in this study}
  \label{PRplotCUReT}
\end{figure}

The results of our experiments on the CUReT database are shown in \Cref{TCUReT}. Precision at one results indicate that the proposed algorithms succeed in retrieving one of the correct matches as a first retrieved image. Precision at two (P@2) results show that the proposed methods are able to retrieve 93\%-94\% of the correct matches in our database. Precision-recall curves for methods based on Gabor features, LBP, default curvelet, curvelet with periodic extension and denoising-based adaptive curvelets are shown in \Cref{PRplotCUReT}. 

\begin{figure}[tb]
  \centering
  \centerline{\includegraphics[width=7cm]{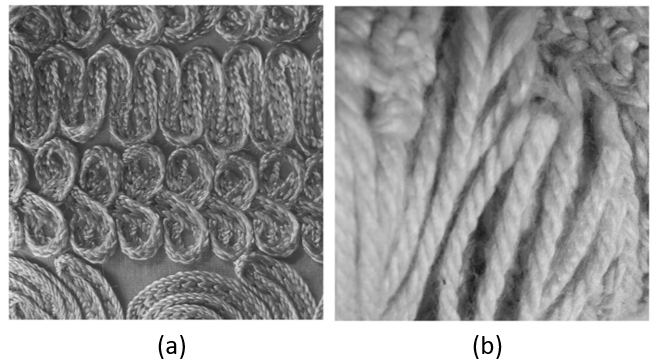}}
  \caption{Curvelet variants failed to retrieve correct matches for these two textural patterns from the Stex-fabric dataset }
  \label{miss}
\end{figure}

Results for our experiments on the fabric database are summarized in \Cref{Tfabric}. The proposed curvelet-based methods and MRELBP outperform other algorithms in the precision of the first retrieved result (P@1). The highest MRR and MAP values in this experiment were achieved by adaptive curvelets. Curvelets with periodic extension and adaptive curvelets successfully retrieved 75 correct matches out of 77 texture classes used in this experiment. The two classes that were misclassified are shown in \Cref{miss}. In contrast to other Stex-fabric images, (a sample of which was shown in \Cref{STex}), these two texture classes do not show uniform and repeated textural patterns. The upper part in image (a) is different from the lower part. In the center of this image, the two rows of textured material are not similar to each other in structure. Similarly, textures are not uniform in image (b). Textures in the upper right corner in image (b) are different from the rest of the image. Moreover, another unique textural area appears in the upper left corner of image (b). 

Adaptive curvelet results can be further improved by training over a set of images resembling the class of interest. The training procedure as it is currently implemented uses the query image. However, training on the query image alone can overemphasize its local features. 

\begin{figure}[tb]
  \centering
  \centerline{\includegraphics[width=7cm]{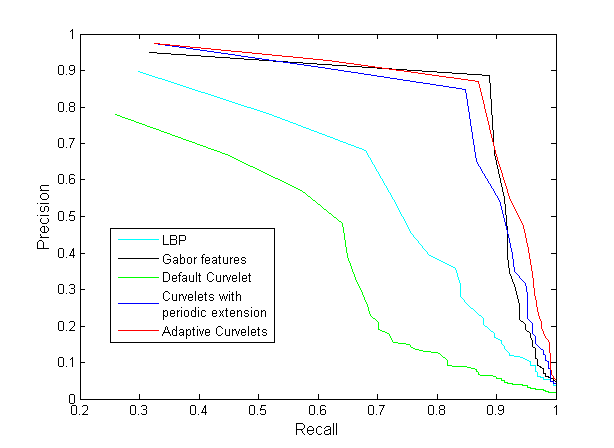}}
  \caption{Precision-recall curves obtained for the Fabric database}
  \label{PRplotfabric}
\end{figure}

Precision-recall curves for methods based on wavelets, Gabor features, default curvelet, curvelet with periodic extension, and denoising-based adaptive curvelets are shown in \Cref{PRplotfabric}.
\begin{table}[tb]
\begin{center}
 \caption{Retrieval results for the Fabric dataset. }
  \begin{tabular}  { |c|c|c|c|c|c| }
    \hline
     & P@1 & P@2 &P@3& MRR  & MAP \\ \hline
        
    Wavelet features &    0.935   &   0.844&0.792 &   0.959    &  0.842   \\ \hline

   Gabor features &  0.948 &0.916 &\textbf{ 0.887 }& 0.961&    0.907 \\ \hline
   Linear Binary Patterns (LBP) &   0.896  &   0.779  & 0.680   &  0.928 &    0.760 \\ \hline
 Completed LBP (CLBP) &   0.948  &   0.851   & 0.762 &0.965 &    0.817 \\ \hline

  Median Robust Extended LBP (MRELBP) &  \textbf{0.974}  &   \textbf{0.929}    & 0.862 &    0.982& 0.901 \\ \hline

   Default curvelet &    0.779  &  0.669 &   0.571 &  0.837 &0.636 \\ \hline
    Curvelet with periodic extension &  \textbf{ 0.974}   & 0.909   & 0.849& 0.983  &0.896\\ \hline
   Adaptive curvelets & \textbf{ 0.974}&0.927  & 0.870  &   \textbf{ 0.984}   &   \textbf{ 0.918}    \\ \hline
  \end{tabular}\label{Tfabric}
\end{center}
\end{table}

\begin{figure}[tb]
  \centering
  \centerline{\includegraphics[width=8cm]{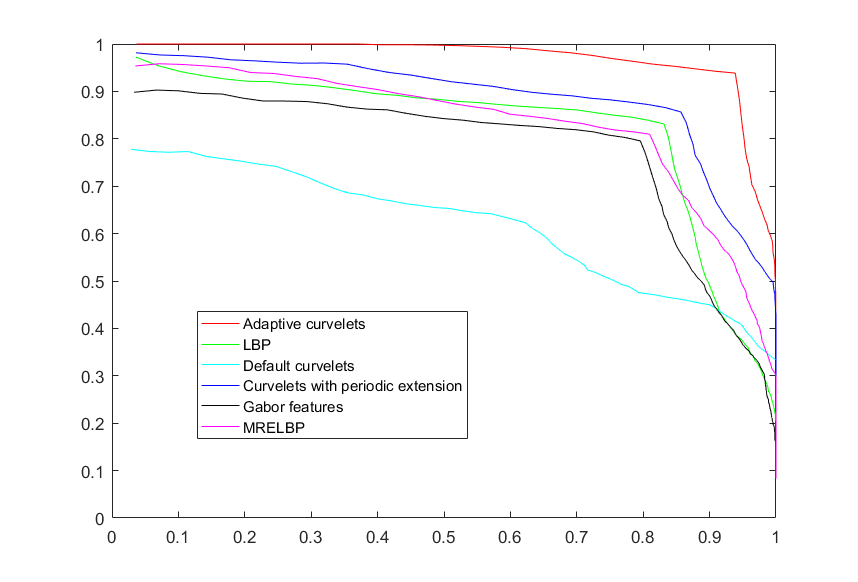}}
  \caption{Precision-recall curves for the Mondial Marmi granite database}
  \label{PRplotgranite}
\end{figure}

Finally, we demonstrate the capabilities of the rotation-normalized version of the algorithm. Results for our experiments on the Mondial Marmi granite database are shown in \Cref{Tgranite}. Adaptive curvelet results are compared with results acquired using LBP, CLBP, MRELBP, and a rotation-normalized version of the Gabor retrieval method. The Gabor retrieval algorithm is normalized to rotations by using the approach detailed in \Cref{rotation-invariant}. Local binary patterns are rotation invariant by construction. The number of divisions per Gabor scale/quadrant pair is kept at the default value of 3. \Cref{Tgranite} presents the results of our experiments, where we compare the performance of adaptive curvelets with LBP variants and the Gabor feature algorithms. It is clear that the rotation normalized adaptive curvelets algorithm outperforms the other algorithms. Adaptive curvelets achieve perfect retrieval precision up to a recall value of 0.5. Precision-recall curves are shown in \Cref{PRplotgranite}.

\begin{table}[tb]
\begin{center}
 \caption{Retrieval results for the Mondial Marmi granite database. Gabor and curvelet-based algorithms were normalized to rotations. }
  \begin{tabular} { |c|c|c|c|c|c| }
    \hline
     & P@1 & P@2 &P@27& MRR  & MAP \\ \hline
   Gabor features &   0.898  &  0.903  &  0.796   & 0.937 &   0.826 \\ \hline
   Linear Binary Patterns (LBP) &    0.972  &  0.954   & 0.831   & 0.973  &  0.880 \\ \hline
  Completed LBP (CLBP) &  0.944  &   0.940    & 0.856 & 0.968 &  0.887 \\ \hline
  Median Robust Extended LBP (MRELBP) & 0.954  &    0.958   &  0.810 & 0.966&  0.868 \\ \hline
   Default curvelets &   0.778 &   0.773  &  0.623 & 0.826& 0.704 \\ \hline
   Default curvelet with periodic extension &   0.982 &   0.977 &   0.857 &   0.989 &   0.907 \\ \hline
   Adaptive curvelets & \textbf{1}&    \textbf{1}  & \textbf{0.939}  &   \textbf{1}   &   \textbf{0.973}    \\ \hline
  \end{tabular}\label{Tgranite}
\end{center}
\end{table}

\subsection{Computational time and feature length}
In this section, we present a comparison between the texture retrieval methods tested in this paper, in running time and feature vector length. \Cref{timelength} lists the different methods used together with the time needed to run the CUReT texture retrieval experiments. In the CUReT experiments 61 images of size 128 by 128 were matched with 122 reference images. There experiments were performed using a laptop equipped with dual 2.5GHz Intel i7 processors and 8GB of RAM. The table also lists the length of feature vectors used in each texture retrieval algorithm. 

Curvelets with periodic extension take a reasonable amount of time and use a reasonable feature vector length. Adaptive curvelets achieved the best retrieval results in all of our experiments. However, learning the optimal curvelet tiling is time consuming. Feature vector length in adaptive and periodically extended curvelets is shorter than MRELBP. Retrieval results of adaptive curvelets and periodically extended curvelets are at least at the level of MRELBP. 

\begin{table}[tb]
\begin{center}
 \caption{Computational time and feature length comparison. Time measures the time needed to complete the CUReT retrieval experiment}
  \begin{tabular} { |c|c|c| }
    \hline
     &  Time (sec.) & Length of feature vector \\ 
     \hline
   Wavelets & 543.2  & 9  \\ \hline 
   Gabor features &   163  &  1152 \\ \hline
   Linear Binary Patterns (LBP) &    322  &  36  \\ \hline
  Completed LBP (CLBP) &  112.7  &   200    \\ \hline
  Median Robust Extended LBP (MRELBP) &  929.8   &   400    \\ \hline
   Default curvelets &   178.8  &   149   \\ \hline
   Default curvelet with periodic extension &   267.2  &   242  \\ \hline
   Adaptive curvelets & 4693.6  & 376 (average)  \\ \hline
  \end{tabular}\label{timelength}
\end{center}
\end{table}

\subsection{Seismic activity characterization}
\begin{figure}[tb]
  \centering
  \centerline{\includegraphics[width=9cm]{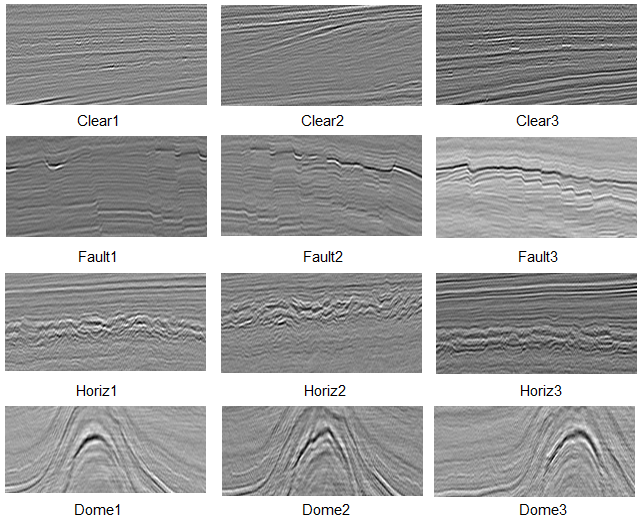}}
  \caption{Seismic images representing different classes of seismic activities.}
  \label{seismicdateset}
\end{figure}

Seismic data are used extensively in the oil and gas exploration industries. Efficient retrieval algorithms based on visual properties of the seismic datasets are desired. The adaptive curvelets retrieval algorithm is tested on the seismic dataset shown in \Cref{seismicdateset}. The figure covers the following seismic activities:
\begin{description}
    \item[Fault:] Datasets with faults
  \item[Horizontal:] Datasets with obvious horizons
  \item[Dome:] Datasets with salt dome shapes and alike
\item[Clear:] Datasets with none of the above activities
\end{description}

\begin{figure}[tb]
  \centering
  \centerline{\includegraphics[width=4cm]{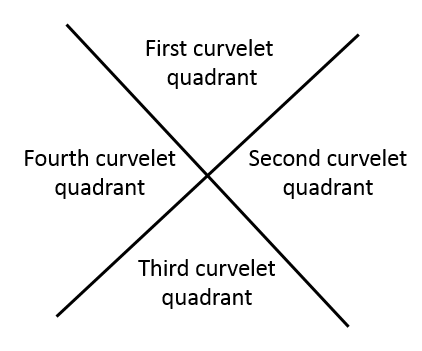}}
  \caption{Curvelet quadrants}
  \label{Curveletquadrants}
\end{figure}

For efficient retrieval performance, a weighted version of the adaptive curvelet algorithm is used. The number of curvelet scale decompositions is chosen, (by the scale selection algorithm), to equal four. \texttt{Fault} images were retrieved using only curvelet wedges representing vertical activities . These wedges lie in the second curvelet quadrant (\Cref{Curveletquadrants}). Only the finer (outer) two curvelet scales are used for retrieval. Weights for the inner scales were set to zero. Similarly, \texttt{Horizontal} datasets were retrieved using curvelet wedges representing horizontal activities (first curvelet quadrant). The inner two curvelet scales (coarsest levels) were used for retrieval. \texttt{Dome} images were retrieved using the outer two curvelet scale and all curvelet directions. \texttt{Clear} images were retrieved using all curvelet scales and directions.

In a previous paper \cite{CurveletCv}, we have shown that for seismic data maximizing the coefficient of variations is an effective method for learning curvelet tiles. We use this alternate cost function to improve default curvelet tiles in this experiment.

Every image in the dataset was used once as a query image, while the remaining images were used as testing images. A comparison between unweighted and weighted versions of adaptive curvelets retrieval performance is shown in \Cref{Tseismic}. Unweighted adaptive curvelets succeed in retrieving the first correct match in 9 out of 12 images in the database. Weighted curvelets increase this ratio to 11 out of 12, where the algorithm matches image \texttt{Horiz3} with \texttt{Clear3} and \texttt{Clear1} instead of the correct \texttt{Horiz} images.

\begin{table}[tb]
\begin{center}
 \caption{Retrieval results for the seismic dataset}
  \begin{tabular} { |c|c|c|c|c|c }
    \hline
     & P@1 & P@2 & MRR  & MAP  \\ \hline
   Adaptive curvelets &   0.75 &   0.75  &   0.85  &  0.83 \\ \hline
   Weighted adaptive curvelets &    \textbf{0.92}  &  \textbf{0.83}    &     \textbf{0.93} &   \textbf{0.91}\\ \hline
  \end{tabular}\label{Tseismic}
\end{center}
\end{table}

\section{Conclusions}
New lightweight algorithms for texture-image retrieval were presented in this paper. Their performance was tested on several texture datasets. Our results strongly encourage the use of periodically extended curvelets, despite the increase in computational cost. Adaptive curvelets were shown to improve default curvelet performance by adapting curvelet tiles to represent the image of interest more efficiently. Adaptive curvelets achieved promising results in retrieving seismic data. In the future, we plan to extend these results by applying the developed texture-retrieval method to video applications. 

\bibliographystyle{elsarticle-num}
\bibliography{main}

\begin{thebibliography}{10}
\expandafter\ifx\csname url\endcsname\relax
  \def\url#1{\texttt{#1}}\fi
\expandafter\ifx\csname urlprefix\endcsname\relax\def\urlprefix{URL }\fi
\expandafter\ifx\csname href\endcsname\relax
  \def\href#1#2{#2} \def\path#1{#1}\fi

\bibitem{Wavelet}
M.~Do, M.~Vetterli, Wavelet-based texture retrieval using generalized gaussian
  density and kullback-leibler distance, Image Processing, IEEE Transactions on
  11~(2) (2002) 146--158.
\newblock \href {http://dx.doi.org/10.1109/83.982822}
  {\path{doi:10.1109/83.982822}}.

\bibitem{Copula}
N.-E. Lasmar, Y.~Berthoumieu, Gaussian copula multivariate modeling for texture
  image retrieval using wavelet transforms, Image Processing, IEEE Transactions
  on 23~(5) (2014) 2246--2261.
\newblock \href {http://dx.doi.org/10.1109/TIP.2014.2313232}
  {\path{doi:10.1109/TIP.2014.2313232}}.

\bibitem{Wavelet2}
Y.~Dong, J.~Ma, Wavelet-based image texture classification using local energy
  histograms, Signal Processing Letters, IEEE 18~(4) (2011) 247--250.
\newblock \href {http://dx.doi.org/10.1109/LSP.2011.2111369}
  {\path{doi:10.1109/LSP.2011.2111369}}.

\bibitem{Wavelet3}
Z.-Z. Wang, J.~Yong, Texture analysis and classification with linear regression
  model based on wavelet transform, Image Processing, IEEE Transactions on
  17~(8) (2008) 1421--1430.
\newblock \href {http://dx.doi.org/10.1109/TIP.2008.926150}
  {\path{doi:10.1109/TIP.2008.926150}}.

\bibitem{Steerable}
E.~Simoncelli, W.~Freeman, E.~Adelson, D.~Heeger, Shiftable multiscale
  transforms, Information Theory, IEEE Transactions on 38~(2) (1992) 587--607.
\newblock \href {http://dx.doi.org/10.1109/18.119725}
  {\path{doi:10.1109/18.119725}}.

\bibitem{Gabor}
T.~S. Lee, Image representation using {2D Gabor} wavelets, Pattern Analysis and
  Machine Intelligence, IEEE Transactions on 18~(10) (1996) 959--971.
\newblock \href {http://dx.doi.org/10.1109/34.541406}
  {\path{doi:10.1109/34.541406}}.

\bibitem{Curvelet}
E.~Candes, L.~Demanet, D.~Donoho, L.~Ying, Fast discrete curvelet transforms,
  Multiscale Modeling {\&} Simulation 5~(3) (2006) 861--899.

\bibitem{Structural}
J.~Zujovic, T.~N. Pappas, D.~L. Neuhoff, Structural texture similarity metrics
  for image analysis and retrieval, Image Processing, IEEE Transactions on
  22~(7) (2013) 2545--2558.

\bibitem{Gabor-texture}
B.~S. Manjunath, W.-Y. Ma, Texture features for browsing and retrieval of image
  data, Pattern Analysis and Machine Intelligence, IEEE Transactions on 18~(8)
  (1996) 837--842.

\bibitem{MPEG-7}
Y.~M. Ro, M.~Kim, H.~K. Kang, B.~Manjunath, J.~Kim, {MPEG-7} homogeneous
  texture descriptor, ETRI journal 23~(2) (2001) 41--51.

\bibitem{curvelet-texture}
D.~Zhang, M.~M. Islam, G.~Lu, I.~J. Sumana, Rotation invariant curvelet
  features for region based image retrieval, International journal of computer
  vision 98~(2) (2012) 187--201.

\bibitem{LBP}
T.~Ojala, M.~Pietik{\"a}inen, D.~Harwood, A comparative study of texture
  measures with classification based on featured distributions, Pattern
  recognition 29~(1) (1996) 51--59.

\bibitem{DOT1}
S.~Hinterstoisser, V.~Lepetit, S.~Ilic, P.~Fua, N.~Navab, Dominant orientation
  templates for real-time detection of texture-less objects, in: 2010 IEEE
  Computer Society Conference on Computer Vision and Pattern Recognition, IEEE,
  2010, pp. 2257--2264.

\bibitem{DOT2}
C.~Hong, J.~Zhu, J.~Yu, J.~Cheng, X.~Chen, Realtime and robust object matching
  with a large number of templates, Multimedia Tools and Applications 75~(3)
  (2016) 1459--1480.

\bibitem{guo2010completed}
Z.~Guo, L.~Zhang, D.~Zhang, A completed modeling of local binary pattern
  operator for texture classification, IEEE Transactions on Image Processing
  19~(6) (2010) 1657--1663.

\bibitem{liu2015median}
L.~Liu, P.~Fieguth, M.~Pietik{\"a}inen, S.~Lao, Median robust extended local
  binary pattern for texture classification, in: Image Processing (ICIP), IEEE
  International Conference on, 2015, pp. 2319--2323.

\bibitem{hu2016completed}
Y.~Hu, Z.~Long, G.~AlRegib, Completed local derivative pattern for rotation
  invariant texture classification, in: Image Processing (ICIP), IEEE
  International Conference on, 2016, pp. 3548--3552.

\bibitem{hu2017scale}
Y.~Hu, Z.~Long, G.~AlRegib, Scale selective extended local binary pattern for
  texture classification, in: Acoustics, Speech and Signal Processing (ICASSP),
  IEEE International Conference on, 2017, pp. 1413--1417.

\bibitem{LTP}
S.~Murala, R.~Maheshwari, R.~Balasubramanian, Local tetra patterns: a new
  feature descriptor for content-based image retrieval, Image Processing, IEEE
  Transactions on 21~(5) (2012) 2874--2886.

\bibitem{CLBC}
Y.~Zhao, D.~Huang, W.~Jia, Completed local binary count for rotation invariant
  texture classification, IEEE Transactions on Image Processing 21~(10) (2012)
  4492--4497.
\newblock \href {http://dx.doi.org/10.1109/TIP.2012.2204271}
  {\path{doi:10.1109/TIP.2012.2204271}}.

\bibitem{SSLBP}
Z.~Guo, X.~Wang, J.~Zhou, J.~You, Robust texture image representation by scale
  selective local binary patterns, IEEE Transactions on Image Processing 25~(2)
  (2016) 687--699.
\newblock \href {http://dx.doi.org/10.1109/TIP.2015.2507408}
  {\path{doi:10.1109/TIP.2015.2507408}}.

\bibitem{long2018comparative}
Z.~Long, Y.~Alaudah, M.~A. Qureshi, Y.~Hu, Z.~Wang, M.~Alfarraj, G.~AlRegib,
  A.~Amin, M.~Deriche, S.~Al-Dharrab, et~al., A comparative study of texture
  attributes for characterizing subsurface structures in seismic volumes,
  Interpretation 6~(4) (2018) T1055--T1066.

\bibitem{liu2017local}
L.~Liu, P.~Fieguth, Y.~Guo, X.~Wang, M.~Pietik{\"a}inen, Local binary features
  for texture classification: taxonomy and experimental study, Pattern
  Recognition 62 (2017) 135--160.

\bibitem{cimpoi2015deep}
M.~Cimpoi, S.~Maji, A.~Vedaldi, Deep filter banks for texture recognition and
  segmentation, in: Computer Vision and Pattern Recognition (CVPR), 2015 IEEE
  Conference on, IEEE, 2015, pp. 3828--3836.

\bibitem{yu2015learning}
J.~Yu, D.~Tao, M.~Wang, Y.~Rui, Learning to rank using user clicks and visual
  features for image retrieval, IEEE transactions on cybernetics 45~(4) (2015)
  767--779.

\bibitem{yu2017deep}
J.~Yu, X.~Yang, F.~Gao, D.~Tao, Deep multimodal distance metric learning using
  click constraints for image ranking, IEEE transactions on cybernetics 47~(12)
  (2017) 4014--4024.

\bibitem{al2017curvelet}
H.~Al-Marzouqi, G.~AlRegib, Curvelet transform with learning-based tiling,
  Signal Processing: Image Communication 53 (2017) 24--39.

\bibitem{aldiss}
H.~Al-Marzouqi, Curvelet transform with adaptive tiling, Dissertation, Georgia
  Institute of Technology (2009).

\bibitem{CurveletGlobalSIP14}
H.~Al-Marzouqi, G.~AlRegib, Texture similarity using periodically extended and
  adaptive curvelets, Global Conference on Signal and Information Processing
  (GlobalSIP), 2014 IEEE.

\bibitem{CurveletSPIE}
H.~Al-Marzouqi, G.~AlRegib, Curvelet transform with adaptive tiling, in: Image
  Processing: Algorithms and Systems X; and Parallel Processing for Imaging
  Applications II, Vol. 8295, International Society for Optics and Photonics,
  2012, p. 82950F.

\bibitem{conn2009introduction}
A.~A.~R. Conn, K.~Scheinberg, L.~N. Vicente, Introduction to derivative-free
  optimization, Vol.~8, {SIAM}, 2009.

\bibitem{RotationFFT}
H.~Bulow, A.~Birk, Spectral {6DOF} registration of noisy {3D} range data with
  partial overlap, Pattern Analysis and Machine Intelligence, IEEE Transactions
  on 35~(4) (2013) 954--969.
\newblock \href {http://dx.doi.org/10.1109/TPAMI.2012.173}
  {\path{doi:10.1109/TPAMI.2012.173}}.

\bibitem{CUReT}
K.~Dana, B.~Van-Ginneken, S.~Nayar, J.~Koenderink, {R}eflectance and {T}exture
  of {R}eal {W}orld {S}urfaces, ACM Transactions on Graphics (TOG) 18~(1)
  (1999) 1--34.

\bibitem{granite}
F.~Bianconi, E.~Gonz{\'a}lez, A.~Fern{\'a}ndez, S.~A. Saetta, Automatic
  classification of granite tiles through colour and texture features, Expert
  Systems with Applications 39~(12) (2012) 11212--11218.

\bibitem{STex}
\href{http://wavelab.at/sources/STex/}{Salzburg texture image database
  ({STex})} (2009).
\newline\urlprefix\url{http://wavelab.at/sources/STex/}

\bibitem{CurveletCv}
H.~Al-Marzouqi, G.~AlRegib, Using the coefficient of variation to improve the
  sparsity of seismic data, in: 2013 IEEE Global Conference on Signal and
  Information Processing, IEEE, 2013, pp. 630--630.

\end{thebibliography}





\end{document}